# Significant Reduction of Graphene Thermal Conductivity by Phononic Crystal Structure

Lina Yang[1], Jie Chen[1,2], Nuo Yang[3*], and Baowen Li[1,4,5,6*]

[1]Department of Physics and Centre for Computational Science and Engineering, National University of Singapore, Singapore 117542, Singapore

[2]Computational Science and Engineering Laboratory, Department of Mechanical and Process Engineering, ETH Zurich, CH-8092 Zurich, Switzerland

[3]School of Energy and Power Engineering, Huazhong University of Science and Technology (HUST), Wuhan 430074, People's Republic of China

[4]Graphene Research Center, National University of Singapore, Singapore 117546, Singapore

[5]NUS Graduate School for Integrative Sciences and Engineering, National University of Singapore, Singapore 117456, Singapore

[6]Center for Phononics and Thermal Energy Science, School of Physics Science and Engineering, Tongji University, 200092 Shanghai, People's Republic of China

*Corresponding authors: N.Y. (nuo@hust.edu.cn) and B.L. (phylibw@nus.edu.sg)

**Abstract**

We studied the thermal conductivity of graphene phononic crystal (GPnC), also named as graphene nanomesh, by molecular dynamics simulations. The dependences of thermal conductivity of GPnCs ($\kappa_{GPnC}$) on both length and temperature are investigated. It is found that the thermal conductivity of GPnCs is significantly lower than that of graphene ($\kappa_G$) and can be efficiently tuned by changing the porosity and period length. For example, the ratio $\kappa_{GPnC}/\kappa_G$ can be changed from 0.1 to 0.01 when the porosity is changed from about 21% to 65%. The phonon participation ratio spectra reveal that more phonon modes are localized in GPnCs with larger porosity. Our results suggest that creating GPnCs is a valuable method to efficiently manipulate the thermal conductivity of graphene.

1. Introduction

As collective lattice vibrations, phonons are responsible for the heat conduction in semiconductors and dielectric materials. Management of phonons provide advances in thermal devices, such as thermal diodes which could control heat flow in a preferred direction,[1-3] thermal cloaking which could hide object from heat,[4-7] thermocrystals which could manipulate heat by phonons,[8] and thermoelectrics which could convert heat into electricity directly or be used as Peltier refrigerator.[9-11]

The graphene has been widely studied because of its fascinating properties. It has extremely high thermal conductivity,[12,13] which is very useful in heat dissipation of electronic devices. Moreover, because contribution to thermal conductivity mainly comes from phonons[12-14], the thermal conductivity of graphene can be manipulated by phonon engineering, such as the size confinement,[15,16] defect scatterings,[17,18] boundary and junctions,[19,20] inter-layer couplings,[21-23] foldings[24,25], mass loading[26] and phononic crystals.[27-29] Therefore, graphene is a promising candidate for phononic engineering devices.

When the period length decreases to nanometers, the transport of terahertz lattice vibrations – phonons in phononic crystals (PnCs) will be affected.[30-36] Due to the periodic change of the density and elastic constant, PnCs could exhibit phononic band gaps at high frequencies.[37] PnCs could also have boundary scatterings[30] and phonon localizations[31] which could block high-frequency phonons. This remarkable property is very different from those of traditional materials, and allows PnCs to achieve new functionalities. There has been a growing attention in recent years to utilize the unique properties of PnCs for thermal applications. It is reported that silicon nanomesh,[38] silicon by PnCs patterning,[39] nanoporous silicon[40] and isotopic silicon PnCs[37] have very low thermal conductivity, which is favorable for thermoelectric applications.

Recently, a graphene PnC (GPnC) also named as nanomesh, a single layer graphene with periodic holes, has been fabricated.[27-29] It is reported that GPnCs could open up a conduction bandgap in graphene and the on-off ratio of the GPnC as a field-effect transistor can be tuned by varying the neck width.[27] Theoretical study has shown that Bragg scatterings lead to a dramatic change in the thermal conductivity of GPnCs, $\kappa_{GPnC}$, considering the very small feature size (7.5 nm in period length).[41] However, the experimental fabrication of graphene with such small period is still challenging. The neck width observed in the previous experimental study of nanomesh graphene[27] is 5~15 nm, corresponding to the period length about 25~75 nm with porosity of 50%.

In this work, using nonequilibrium molecular dynamics (NEMD) method, we study the thermal conductivity of GPnCs with period length ($L_0$) varying from 10.4 nm to 62.5 nm. The dependences of thermal conductivity on both length ($L$) and temperature ($T$) are studied. Moreover, the porosity is tuned from 0.0041% to 66% by varying the diameter ($D$) of holes, and its impact on the thermal conductivity of GPnCs is investigated. Vibrational eigen-modes analysis reveals that phonon modes localization in GPnCs is enhanced at large porosity.

2. Structure and method

The structure of GPnC is a single layer graphene embedded with periodic circular holes, which is characterized by length $L_0$ and diameter $D$. Neck width corresponds to $L_0$–$D$. The same $L_0$ is used in the longitudinal and transverse direction. Fig. 1(a) shows a simulation cell with 5 periods in longitudinal direction and 1 period in transverse direction, where $L_0$=25 nm and $D$=15 nm.

The periodic boundary condition is used in transverse direction and the fixed boundary condition is used in longitudinal direction. Optimized Tersoff potential is applied to describe the C-C interactions, which can better describe the lattice properties of graphene.[42] A temperature gradient is established along the longitudinal direction by applying Langevin heat bath[43] at the two ends (boxes in Fig. 1(a)).

The equations of motions are integrated by velocity Verlet method with a time step of 0.5 fs. In the beginning, the simulation runs 0.6 ns to reach a steady state by applying heat bath in an NVT ensemble (constant number of atoms, volume, and $T$). Then the simulation runs 7 ns to get an averaged heat current and temperature profile. The thermal conductivity is calculated from Fourier's law

$$\kappa = -\frac{J \cdot L}{A \cdot \Delta T} \qquad (1)$$

where $J$ is the heat current, $L$ is the length of the simulation cell, $\Delta T$ is the temperature difference and $A$ is the cross section area. $A = W \times 3.4\,\text{Å}$, where $W$ is the width and 3.4 Å is the graphene inter-layer distance.[17,44,45] The final result is the mean value of twelve realizations with different initial conditions. The error bar is the standard deviations of the results of 12 simulations. As the heat transport in graphene is isotropic, the thermal conductivity is the same in zigzag and armchair direction.[21,46,47] Only zigzag graphene and zigzag GPnCs are studied in this work.

In Fig. 1(b), we show the temperature profile of a simulation cell. The temperatures at two ends are set as $T_L$=310 K and $T_R$=290 K. There are temperature jumps at the two boundaries due to the couplings with heat baths. The $\Delta T$ is defined as the temperature difference between the two dash lines, which excludes the temperature jumps next to heat bath.

3. Simulation results

The size effect could arise if the simulation cell is not sufficiently large in the transverse direction.[48] Based on the period length $L_0$=25 nm and length $L$=125 nm, we examined the width effect of simulation cell on the thermal conductivity. The room temperature thermal conductivities of GPnC with the width one period and two periods are 143.05±2.84 and 143.09±1.33 W/m-K, respectively. That is, one period in transverse direction is enough to get a saturated thermal conductivity. In the following simulations, we choose one period in transverse direction in all simulation cells of GPnCs.

Along the heat transfer direction (longitudinal), the thermal conductivity depends on the length of material in nanoscale. Previous works[14,21,46] on graphene found that the thermal conductivity increases as the length of graphene increases and the thermal conductivity does not converge even at several micrometers. We study the $\kappa_{GPnC}$ dependence on the length of GPnCs at room temperature, which is shown in Fig. 2(a). Inset zooms in for GPnC at the small scale. Here, the $L_0$ is fixed as 25 nm, and $D$ is fixed as 15 nm. With the increase of length, thermal conductivity first increases, and then increases slowly after $L \geqslant 250$ nm.

In comparison with GPnCs, Fig. 2(a) also shows the length dependence of $\kappa_G$ at 300 K. The same boundary conditions and heat bath as in the GPnCs are used. The width of graphene simulation cell is set as 5.2 nm, which is large enough to simulate the thermal conductivity of graphene.[46] Our simulation result of $\kappa_G$ with $L$=300 nm is 1694±37 W/m-K, which is consistent with previous works.[21,46] The red dash line is proportional to log($L$), which indicates the $\kappa_G$ follows ~ log($L$) behavior, which has been found in the experimental study.[14] Moreover, we found $\kappa_G$ does not converge to a finite value within 300 nm in length. The effective phonon mean free path in graphene is 240 nm at 300 K, which may lead to the significant length dependence and high thermal conductivity.[14]

Compared with graphene, the GPnCs have much lower values of thermal conductivity as shown in Fig. 2(a), and the phonon dispersion curves of GPnC are folded and flattened (details in Fig. 4(b) and (c)). There should be more phonon-phonon couplings in GPnCs. The phonons participating in Umklapp three-phonon scattering should satisfy the momentum and energy conservation as $q \pm q' = q'' + K$, where $K$ is a reciprocal lattice vector and $\omega(q) \pm \omega(q') = \omega(q'')$.[45,49] We counted the number of phonons states available for Umklapp three-phonon scattering ($N_{sc}$) and found that $N_{sc}$ in GPnCs (same as in Fig. 4(c)) increases by a factor of 127 as compared with graphene. That is, in the GPnCs, there are much more states available for Umklapp three-phonon scatterings which could limit the thermal conductivity.[24,49] In addition, the phonon group velocities are reduced due to the flattened phonon dispersion curves, and phonons are more likely localized in GPnCs (details in Fig. 4(a)). Therefore, the thermal conductivity of GPnCs are much lower than that of graphene.

The temperature dependence of thermal conductivity of graphene has been studied in previous works.[46,50,51] In Fig. 2(b), we compared the temperature dependence of $\kappa_G$ with that of $\kappa_{GPnC}$. For GPnCs, we fixed $L_0$ as 25 nm, $D$ as 15 nm and $L$ as 125 nm. The graphene has the same $L$ as the GPnCs. The dash line draws the $1/T$ temperature scaling for reference. The $\kappa_G$ with $L$=125 nm approximately decreases in proportion to $1/T$, which comes from the Umklapp phonon-phonon scattering. In the range from 300 K to 500 K, the $\kappa_G$ with $L$=125 nm decreases slightly faster than the $1/T$ curve, which is caused by the second order three phonon scattering process.[46,52] The $\kappa_{GPnC}$ decreases more slowly than $\kappa_G$ as the temperature increases from 300 K to 800 K.

Besides the length dependence, the diameter of holes could be another factor affecting the thermal conductivity. We fixed $L_0$ as 25 nm and $L$ as 125 nm, varied $D$ from 0.26 nm to 23 nm. The porosity is defined as the ratio of the number of removed C atoms to the total number of C atoms before removal in one period. The calculation result for $\kappa_G$ with $L$=125 nm is 1392±29 W/m-K at room temperature (Fig. 2(a)). The ratio $r$ of $\kappa_{GPnC}$ to $\kappa_G$ at room temperature is shown in Fig. 3 (a). As the porosity increases from zero to 3%, $r$ drops rapidly from 1.0 to 0.61. Then it slowly decreases to 0.01 when the porosity reaches 66%. Especially, when there is only one atom removed in each period, the porosity of this GPnC is calculated as 0.0041%, and the thermal conductivity is decreased from 1392 (graphene) to 1188 (GPnC) W/m-K. That is the smallest porosity 0.0041% will cause the reduction of thermal conductivity by a factor of 0.146. Previous work found that PnCs with larger porosity has lower phonon group velocities.[30,37] Additionally, phonon modes are more localized in PnC with larger porosity.[31] Consequently, GPnCs with larger porosity will have a smaller thermal conductivity. This result indicates that the thermal conductivity could be effectively tuned by changing the diameter of holes in GPnCs.

Furthermore, we investigate the dependence of thermal conductivity on the period length. With the fixed porosity of 28% and $L$=125 nm, we changed the $L_0$ from 10.4 nm to 62.5 nm. The $\kappa_G$ with $L$=125 nm is calculated as 1392±29 W/m-K at room temperature (Fig. 2(a)). The ratio $r$ of $\kappa_{GPnC}$ to $\kappa_G$ at room temperature is shown in Fig. 3(b). We find $r$ increases as the period length increases. PnC with smaller period length will have denser boundary scatterings. Previous works have found that Umklapp scattering and boundary scatterings dominate in PnCs with large period length.[30,32] As the boundary scatterings become denser, the thermal conductivity decreases monotonically. This result suggests that period length is another important factor that could tune the thermal conductivity.

In order to understand the underlying physical mechanism of the reduction of thermal conductivity, a vibrational eigen-mode analysis on GPnCs is carried out.[37,53] The participation ratio ($P$) for phonon mode k is defined through the normalized eigenvector $u_{i\alpha,k}$

$$P_k = \frac{1}{N\sum_{i=1}^{N}\left(\sum_{\alpha=1}^{3}u_{i\alpha,k}^{*}u_{i\alpha,k}\right)^2} \quad (1)$$

where $N$ is the total number of atoms, $u_{i,\alpha,k}$ is calculated by general utility lattice program (GULP).[54] When there are less atoms participating in the motion, the phonon mode has a smaller value of $P$.

Participation ratios of phonons in graphene and GPnCs with different porosity are computed in Fig. 4(a). As shown in Fig. 4(a), the participation ratios of GPnCs are significantly reduced compared with that of graphene, thus, GPnCs have smaller thermal conductivity than graphene. Additionally, there exists an obvious reduction of the participation ratios in GPnCs when the porosity changes from 10% to 50%, which indicates that larger porosity could enhance phonon localizations in GPnCs. This is responsible for the results in Fig. 3(a) that GPnCs with larger porosity has lower thermal conductivity. Similar results are also found in Si PnCs.[31] Therefore, thermal conductivity can be manipulated by varying the porosity. Fig. 4(b) and (c) shows the lower-frequency part of phonon dispersion of graphene and GPnC, respectively. The phonon dispersions of graphene and GPnC are calculated by lattice dynamics implemented in GULP.[54] Apparently, the phonon dispersions are folded and flattened in GPnC. Based on our calculations, there are much more states available for three-phonon Umklapp scatterings in GPnC, which will cause the reduction of thermal conductivity.

4. Conclusion

We have studied the thermal conductivity of GPnCs by using nonequilibrium molecular dynamics method. The simulation results show that thermal conductivity of GPnCs increases as the length increases, then increases slowly when the length is larger than 250 nm. It is found that the value of the thermal conductivity of GPnCs is significant reduced as compared with graphene. Above room temperature, thermal conductivity of graphene decreases approximately proportional to $1/T$, while thermal conductivity of GPnCs decreases more slowly than that of graphene as the temperature increases. Thermal conductivity of GPnCs is greatly decreased with the increase of porosity and the decrease of period length. Vibrational eigen-modes analysis reveals that the increase of porosity enhances phonon localizations in GPnCs, which lead to the reduction of thermal conductivity. Our results suggest that creating GPnCs is a useful method to efficiently control the thermal conductivity of graphene.

**Acknowledgments**

B.L. was supported in part by the grants from the Asian Office of Aerospace R&D of the US Air Force (AOARD-114018), MOE Grant R-144-000-305-112 of Singapore, and the National Natural Science Foundation of China (11334007). N.Y. was supported in part by the grants from Self-Innovation Foundation (2014TS115) of HUST, Talent Introduction Foundation (0124120053) of HUST, and the National Natural Science Foundation of China Grant (11204216). We are grateful to Sha Liu and Liyan Zhu for useful discussions. The authors thank the National Supercomputing Center in Tianjin (NSCC-TJ) for providing assistance in computations.

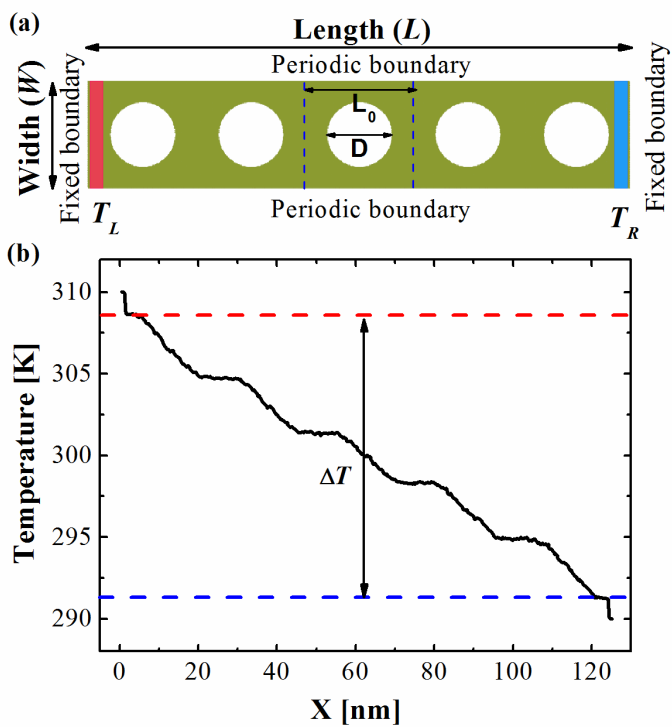

FIG. 1(a). Simulation setup and structure of GPnC with periodic circular holes. $L_0$ is the period length, $L$ is the length of simulation cell, and $D$ is the diameter of the hole. The same $L_0$ is used in the longitudinal and transverse direction. $L_0$=25 nm, $D$=15 nm and $L$=125 nm in (a). Periodic boundary condition is used along transverse direction and fixed boundary condition is used in longitudinal direction. Heat bath with higher temperature (red box) $T_L$ and lower temperature (blue box) $T_R$ are applied at two ends. (b). Temperature profile of simulation of GPnC in (a). Temperature of two heat baths is set as $T_L$=310 K and $T_R$=290 K. There are boundary jumps at the two ends. Temperature difference $\Delta T$ is between the two dash lines.

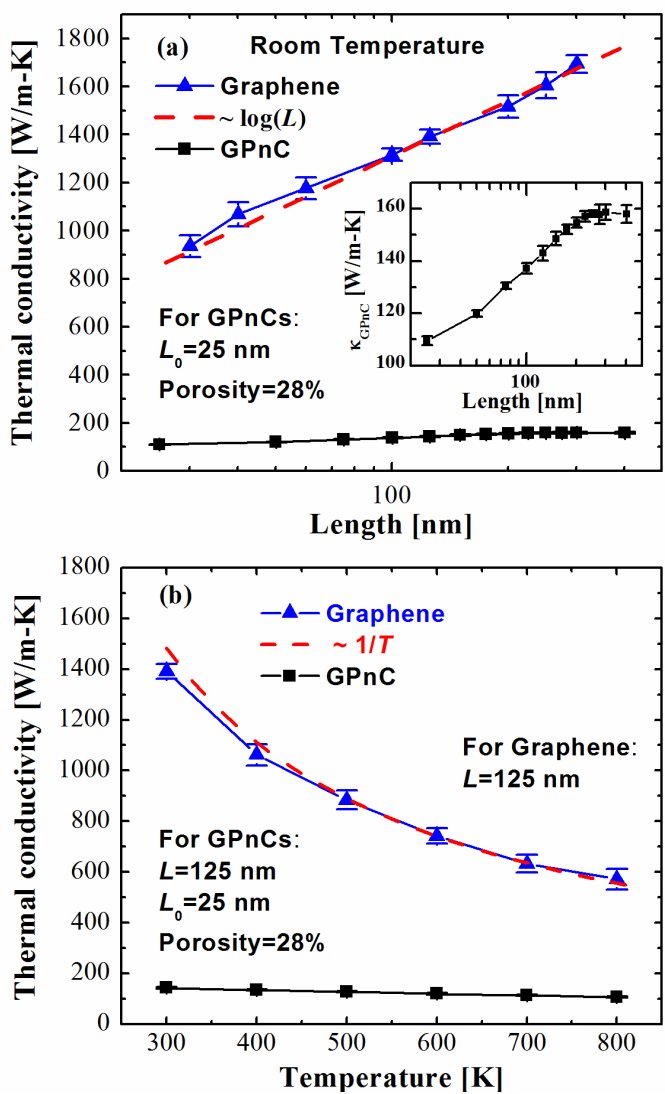

FIG. 2(a). Thermal conductivity of graphene and GPnC versus length of simulation cell at 300 K. The width of simulation cell of graphene is 5.2 nm. For GPnC, $D$=15nm, $L_0$=25 nm and the porosity is 28%. The red dash line is proportional to $\log(L)$, which is used for references. Inset zooms in for GPnC at the small scale. (b). Thermal conductivity of graphene and GPnCs versus temperature. For GPnCs, $L_0$=25 nm, $D$=15 nm and $L$=125 nm. For graphene, $L$ is also 125 nm and the width is 5.2 nm. The red dash line is proportional to $1/T$, which is used for references.

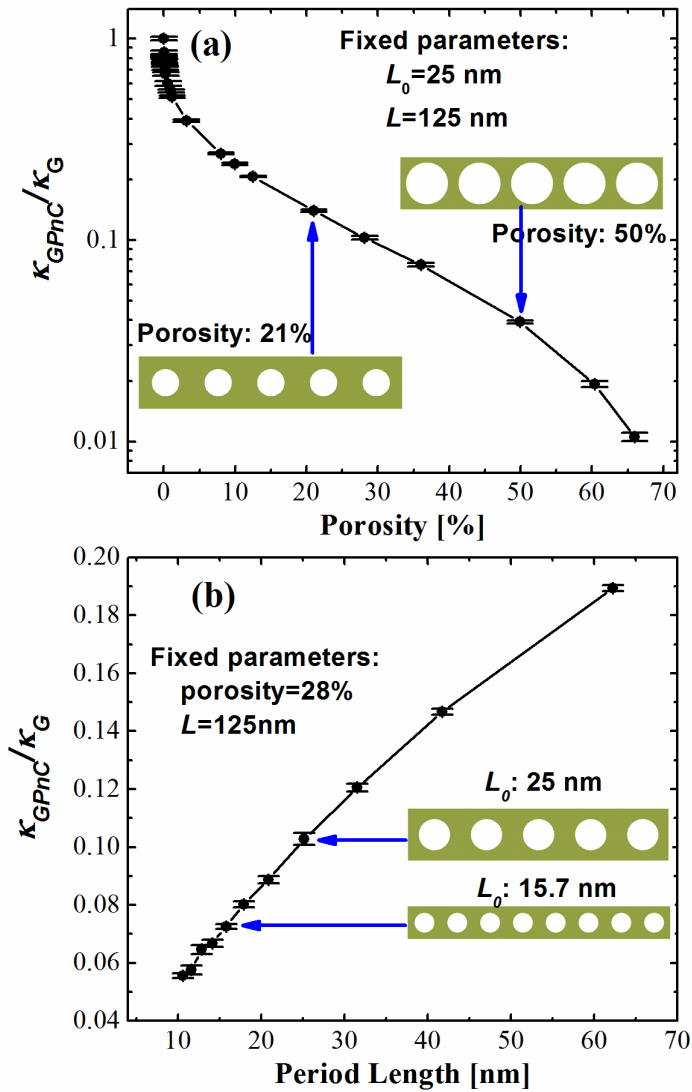

FIG. 3(a). The ratio $r$ of $\kappa_{GPnC}$ to $\kappa_G$ versus porosity at 300 K. For GPnCs, $L$=125nm, $L_0$=25 nm and $D$ vary from 0.26 nm to 23 nm. The $\kappa_G$ with $L$=125 nm is calculated as 1392±29 W/m-K. The insert figure is the simulation cell of GPnCs with porosity 50% and 21% respectively. (b). The ratio $r$ of $\kappa_{GPnC}$ to $\kappa_G$ versus period length at 300 K. For GPnCs, $L$=125 nm, the porosity is fixed as 28% and $L_0$ vary from 10.4 nm to 62.5 nm. The insert figure is the simulation cells of GPnCs with $L_0$=25 nm and 15.7 nm respectively.

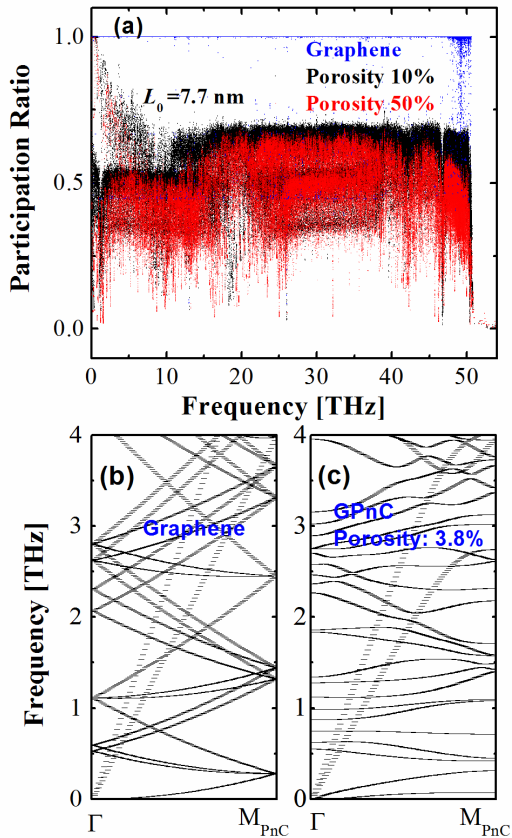

Fig. 4(a) Participation ratio spectra of graphene and GPnCs with different porosity. The $L$ of the square cell of graphene is set as 2.55 nm for the calculation of the participation ratios and phonon dispersion. $L_0$ is fixed as 7.7 nm for GPnCs. Red points and black points correspond to participation ratio of GPnCs with porosity 50% and 10%, respectively. Blue points correspond to participation ratios of graphene. GPnCs has smaller participation ratio than graphene. Additionally, GPnCs with smaller porosity have larger participation ratio. (b) Low frequency part of graphene phonon dispersion. (c) Low frequency part of GPnC phonon dispersion. The $M_{PnC}$ is the M point of square Brillouin zone. The square graphene cell in (b) and the GPnC in (c) are the same as the graphene cell in (a), and the porosity of the GPnC is set 3.8%. Phonon dispersions in GPnC are flattened compared with that in graphene.